\documentclass[twocolumn,superscriptaddress,floats,prd,nofootinbib,showpacs]{revtex4}

\usepackage{amsmath,amssymb,amsfonts}
\usepackage{graphicx}

\def\be{\begin{equation}}
\def\ee{\end{equation}}
\def\ba{\begin{eqnarray}}
\def\ea{\end{eqnarray}}
\def\nn{\nonumber}

\newcommand{\Pl}{\ell_\textrm{Pl}} 
\newcommand{\mubar}{{\bar \mu}} 
\newcommand{\abs}[1]{{\left|{#1}\right|}} 

\newcommand{\secref}[1]{Sec.~\ref{#1}}
\newcommand{\eqnref}[1]{(\ref{#1})}
\newcommand{\figref}[1]{Fig.~\ref{#1}}

\begin{document}


\title{How loopy is the quantum bounce? --- A heuristic analysis\\ of higher order holonomy corrections in LQC}
\author{Dah-Wei Chiou}
\email{chiou@gravity.psu.edu}
\affiliation{
Department of Physics, Beijing Normal University, Beijing 100875, China}
\author{Li-Fang Li}
\email{lilifang@mail.bnu.edu.cn}
\affiliation{
Department of Physics, Beijing Normal University, Beijing 100875, China}

\begin{abstract}
A well-motivated extension of higher order holonomy corrections in loop quantum cosmology (LQC) for the $k=0$ Friedmann-Robertson-Walker model is investigated at the level of heuristic effective dynamics. It reveals that the quantum bounce is generic, regardless of the order of corrections, and the matter density remains finite, bounded from above by an upper bound in the regime of the Planckian density, even if all orders of corrections are included. This observation provides further evidence that the quantum bounce is essentially a consequence of the loopy nature (i.e. intrinsic discreteness) of LQC and LQC is fundamentally different from the Wheeler-DeWitt theory; it also encourages one to construct the quantum theory of LQC with the higher order holonomy corrections, which might be understood as related to the higher $j$ representations in the Hamiltonian operator of loop quantum gravity.
\end{abstract}


\maketitle


\section{Introduction}
Over the last years, the status of loop quantum cosmology (LQC) has progressed significantly and has become an active area of research. Specifically, with the inclusion of a free massless scalar field, the comprehensive formulation of LQC in the $k=0$ Friedmann-Robertson-Walker (FRW) (i.e. spatially flat and isotropic) model has been constructed in detail, giving a solid foundation for the quantum theory and revealing that the big bang singularity is resolved and replaced by the \emph{quantum bounce}, which bridges the present universe with a preexisting one \cite{Ashtekar:2006rx,Ashtekar:2006uz,Ashtekar:2006wn}. Resolution of the classical singularity and occurrence of the quantum bounce have been shown to be robust \cite{Ashtekar:2007em}, and similar results are also affirmed for extended models \cite{Ashtekar:2006es,Vandersloot:2006ws,Chiou:2006qq,Szulc:2008ar}.

However, despite an attractive and long-sought feature, it is questionable whether the quantum bounce results intimately from the quantum nature of Riemannian geometry of loop quantum gravity (LQG), as the same result can be easily obtained even at the level of heuristic effective dynamics without invoking the sophisticated features of LQC \cite{Chiou:2007dn,Chiou:2007sp,Chiou:2007mg}.

To quickly see this, we start with the classical Hamiltonian constraint for the $k=0$ FRW model:\footnote{More precisely, $p$ should be $\abs{p}$ in \eqnref{eqn:classical H}. The sign of $p$ corresponds to spatial orientation, which is irrelevant for our purpose.}
\be\label{eqn:classical H}
H=H_\textrm{grav}+H_\phi
=-\frac{3}{8\pi G\gamma^2}\,c^2\sqrt{p}+\frac{p_\phi^2}{2p^{3/2}},
\ee
where $c$ and $p$ are the Ashtekar variables satisfying the canonical relation
\be\label{eqn:c and p}
\{c,p\}=\frac{8\pi G\gamma}{3}
\ee
with $\gamma$ being the Barbero-Immirzi parameter and $p_\phi$ the conjugate momentum of the free massless scalar field $\phi(\vec{x},t)=\phi(t)$ with
\be
\{\phi,p_\phi\}=1.
\ee

Next, at the heuristic level, we take the prescription of ``holonomization'' to replace $c$ with
\be\label{eqn:holonomization}
c \longrightarrow \frac{\sin(\mubar c)}{\mubar}
\ee
by introducing the discreteness variable $\mubar$. The heuristic effective dynamics is then solved as if the dynamics was classical but governed by the new ``holonomized'' Hamiltonian, which reads as
\be\label{eqn:H mubar}
H_\mubar
=-\frac{3}{8\pi G\gamma^2}\frac{\sin^2\mubar c}{\mubar^2}
\sqrt{p}+\frac{p_\phi^2}{2p^{3/2}}.
\ee
Particularly, in the improved dynamics suggested by \cite{Ashtekar:2006wn}, $\mubar$ is given by
\be\label{eqn:mubar scheme}
\mubar=\sqrt{\frac{\Delta}{p}}\,,
\ee
where $\Delta$ is the area gap in the full theory of LQG and $\Delta=2\sqrt{3}\pi\gamma\Pl^2$ for the standard choice (but other choices are also possible) with $\Pl:=\sqrt{G\hbar}$ being the Planck length.
With \eqnref{eqn:mubar scheme} imposed, the modified Hamiltonian constraint $H_\mubar=0$ immediately sets an upper bound for the matter density:
\be\label{eqn:formal boundedness}
\rho_\phi:=\frac{p_\phi^2}{2p^3}
=\frac{3}{8\pi G\gamma\Delta}\sin^2\mubar c
\leq 3\rho_\textrm{Pl},
\ee
where the Planckian density is defined as
\be
\rho_\textrm{Pl}:=(8\pi G \gamma^2\Delta)^{-1}.
\ee

Apparently, without going into the detailed construction of LQC at all, it is anticipated that the matter density is bounded above and thus the quantum bounce is expected.\footnote{If a different scheme other than \eqnref{eqn:mubar scheme} is adopted, $\rho_\phi$ can also be shown to bounded above, as long as $\mubar$ is prescribed to be $\mubar\propto (\Delta/p)^{r}$ with $r>-1$, but the upper bound depends on the constant of motion $p_\phi$ except the case of $r=1/2$.} One might then argue that the boundedness of $\rho_\phi$ has little to do with the fundamental structure of LQC but merely results from the formal modification of \eqnref{eqn:holonomization}. In this sense, the quantum bounce seems to be an \textit{ad hoc} phenomenon and not really ``loopy'' enough.

In response to this criticism, a simplified but exactly soluble model of LQC has been studied and used to show that the quantum bounce is generic, not just restricted to the states which are semiclassical at late times \cite{Ashtekar:2007em}. Furthermore, the study of \cite{Ashtekar:2007em} brings out the precise sense in which the Wheeler-DeWitt (WDW) theory approximates LQC and the sense in which this approximation fails, thereby showing that LQC is intrinsically discrete and the underlying discreteness is essential for the quantum bounce.

To add further evidence for the loopy nature of the quantum bounce, we explore a new avenue by investigating the well-motivated extension of higher order holonomy corrections, with which the prescription of holonomization is more involved than \eqnref{eqn:holonomization}, yet the quantum bounce is still ensured, at least at the level of heuristic effective dynamics.

\section{Higher order holonomy corrections}\label{sec:holonomy corrections}
One of the very features of LQC is that the connection variable $c$ does not exist and should be replaced by holonomies. Following the standard techniques in gauge theories,
components of the curvature $F=\tau_kF^k_{ij}dx^i\wedge dx^j$ can be expressed in terms of holonomies (i.e. Wilson loops) as
\be\label{eqn:holonomy approx}
F^k_{ij} \approx
-\frac{2}{\mubar^2L^2}\textrm{Tr}
\left[\tau_k \left(h_{\Box_{ij}}^{(\mubar)}-\openone\right)
\right],
\ee
where $2i\tau_k=\sigma_k$ are the Pauli matrices and $h_{\Box_{ij}}^{(\mubar)}$ is the holonomy around the square $\Box_{ij}$ whose edges are parallel to the $i$ and $j$ directions and of coordinate length $\mubar L$ ($L$ is the coordinate length of the edges of the fiducial cell $\mathcal{V}$ prescribed to make sense of the Hamiltonian). This amounts to approximating $c^2$ in \eqnref{eqn:classical H} by ${\sin^2 \mubar c}/{\mubar^2}$
and thus the prescription of \eqnref{eqn:holonomization}. (For more details, see Appendix B of \cite{Chiou:2007mg}.)

In the standard treatment of LQC, instead of shrinking $\Box_{ij}$ to a point, the expression \eqnref{eqn:holonomy approx} is regarded as fundamental and $\mubar$ is set to a nonzero value as an imprint of the discrete area spectrum in the full theory of LQG. It is natural to ask whether the approximation \eqnref{eqn:holonomy approx} can be improved; that is, is it possible to approximate the curvature $F$ in terms of holonomies along edges of finite lengths, yet with arbitrary accuracy?

The approximation \eqnref{eqn:holonomy approx} is based on the Stokes' theorem and, in general, it becomes exact only in the limit when $\Box_{ij}$ shrinks to a point. In the context of cosmologies, however, thanks to homogeneity, for any given finite $\mubar$, it is still possible to approximate $c$ in terms of $\sin\mubar c$ to arbitrary accuracy.
As a heuristic approach, disregarding the standard Stokes' theorem but instead considering the Taylor series
\be
\sin^{-1}x=\sum_{k=0}^\infty \frac{(2k)!}{2^{2k}(k!)^2(2k+1)}\,x^{2k+1}
\ee
for $-1\leq x \leq 1$ and setting $x=\sin \mubar c$, we have
\be
c=\frac{1}{\mubar}\sum_{k=0}^\infty \frac{(2k)!}{2^{2k}(k!)^2(2k+1)}\,
{(\sin \mubar c)}^{2k+1}.
\ee
This inspires us to define the $n$th order holonomized connection variable as
\be\label{eqn:holonomized c}
c_h^{(n)}:=\frac{1}{\mubar}\sum_{k=0}^{n} \frac{(2k)!}{2^{2k}(k!)^2(2k+1)}\,
{(\sin \mubar c)}^{2k+1},
\ee
which can be made arbitrarily close to $c$ (as $n\rightarrow\infty$) but remains a function of the holonomy $\sin \mubar c$ and the discreteness variable $\mubar$. Therefore, to implement the underlying structure of LQC by replacing $c$ with holonomies, $c_h^{(n)}$ can be used as an improved version of \eqnref{eqn:holonomization}, which now reads as $c_h^{(n=0)}$.

Instead of \eqnref{eqn:H mubar}, the Hamiltonian with holonomy corrections up to the $n$th order can be designated as
\be\label{eqn:H mubar n}
H_\mubar^{(n)}
=-\frac{3}{8\pi G\gamma^2}\,(c_h^{(n)})^2
\sqrt{p}+\frac{p_\phi^2}{2p^{3/2}}.
\ee
If we take \eqnref{eqn:H mubar n} as the departing point for LQC, the quantum theory is much more difficult to construct, but it might be possible to treat the higher order corrections as perturbations based on the well-established $n=0$ formalism. The higher order corrections correspond to higher powers of $\sin\mubar c$, which might be understood as the imprint of generic $j$ representations for holonomies in the Hamiltonian operator in the full theory of LQG \cite{Gaul:2000ba,Perez:2005fn}. (See also \cite{Vandersloot:2005kh} for the issues of $j$ ambiguity in LQC.) Thus, even though $c_h^{(n)}$ is obtained heuristically, the modified Hamiltonian of \eqnref{eqn:H mubar n} may reflect the underlying physics of LQG in a more elaborate fashion.

One might suspect that if we include corrections of all orders, the quantum theory of $H_\mubar^{(n=\infty)}$ will lead to the same result of the WDW theory, as $c_h^{(\infty)}=c$ formally. This should not be the case, because the elementary variables are still $\sin \mubar c$ and $p$, instead of $c$ and $p$, even in the limit $n\rightarrow\infty$ and, therefore, the striking difference between LQC and the WDW theory as emphasized in \cite{Ashtekar:2007em} should persist.

While the quantum theory of LQC could be extremely difficult, we will investigate the ramifications of the higher order holonomy corrections at the level of heuristic effective dynamics. Beforehand, using \eqnref{eqn:c and p}, we compute
\ba
\{c,c_h^{(n)}\}
&=&\frac{8\pi G\gamma}{3\mubar}\frac{\partial\mubar}{\partial p}
\left[
\cos(\mubar c)\,\mathfrak{S}_n(\mubar c)\,c-c_h^{(n)}
\right],\quad\\
\{p,c_h^{(n)}\}
&=&-\frac{8\pi G\gamma}{3}\cos(\mubar c)\,\mathfrak{S}_n(\mubar c),
\ea
where
\ba
\mathfrak{S}_n(\mubar c)&:=&
\sum_{k=0}^{n} \frac{(2k)!}{2^{2k}(k!)^2}\,{(\sin \mubar c)}^{2k}\\
&
\mathop{\longrightarrow}\limits_{n \rightarrow\infty}
&
{\abs{\cos \mubar c}}^{-1}.
\ea

\textbf{Remark.} One should not confuse the extension of higher order holonomy corrections discussed here with that studied in \cite{Mielczarek:2008zz} and \cite{Hrycyna:2008yu}. The former extends $c_h^{(n=0)}$ to $c_h^{(n)}$ with the inclusion of higher powers of $\sin\mubar c$, which are most likely to be interpreted as higher $j$ representations, while the latter takes into account the error corrections in \eqnref{eqn:holonomy approx} with respect to the powers of $\mubar$. That is, the former is about the $\mathcal{O}(\sin^{2n+1}\mubar c)$ corrections for $c$ while the latter is about the $\mathcal{O}(\mubar^{2n+2})$ corrections on $\sin^2\mubar c$. These two extensions are motivated differently and give distinct dynamical behaviors. As will be seen, at least for heuristic effective dynamics, the study of this paper affirms that, for any order $n$, the nonsingular bouncing scenario is generic, as opposed to the result of \cite{Mielczarek:2008zz} and \cite{Hrycyna:2008yu}, which indicates that the $\mathcal{O}(\mubar^4)$ corrections lead to the nonperturbative effects (\emph{hyper-inflation}/\emph{deflation}) and thus yield a qualitatively different cosmological scenario in which the bounce and singularity coexist. More comments on the distinction are given in \secref{sec:discussion}.

\section{Heuristic effective dynamics}\label{sec:heuristic dynamics}
At the level of heuristic effective dynamics, the evolution is solved as if the dynamics was classical but governed by the new Hamiltonian \eqnref{eqn:H mubar n} with holonomy corrections up to the $n$th order. For the case of $n=0$, it has been shown that this heuristic treatment gives a very good approximation for the quantum evolution of LQC, and the bouncing scenario of the effective solution gives the absolute upper bound for the matter density \cite{Bojowald:2006gr,Chiou:2008bw}. For $n>0$, the reliability remains to be justified, but the heuristic analysis can still provide good ideas of what the quantum evolution may look like in the presence of higher order holonomy corrections.

By choosing the lapse function $N=p^{3/2}$ associated with the new time variable $t'$ via $dt'=N^{-1}dt$ ($t$ is the proper time), the modified Hamiltonian \eqnref{eqn:H mubar n} is rescaled and simplified as
\be
H_\mubar^{(n)'}=-\frac{3}{8\pi G\gamma^2}\,(c_h^{(n)})^2p^2+\frac{p_\phi^2}{2}.
\ee
The effective equations of motion are then given by Hamilton's equations:
\ba
\label{eqn:eom 1}
\frac{dp_\phi}{dt'}&=&\{p_\phi,H_\mubar^{(n)'}\}=0\quad\Rightarrow\quad
p_\phi\ \text{is constant},\\
\label{eqn:eom 2}
\frac{d\phi}{dt'}&=&\{\phi,H_\mubar^{(n)'}\}=p_\phi,\\
\label{eqn:eom 3}
\frac{dc}{dt'}&=&\{c,H_\mubar^{(n)'}\}=-\frac{2}{\gamma}(c_h^{(n)})^2p\\
&&\qquad -\frac{2}{\gamma}\frac{1}{\mubar}\frac{\partial \mubar}{\partial p}
\left[\cos(\mubar c)\,\mathfrak{S}_n(\mubar c)\,c-c_h^{(n)}\right]c_h^{(n)}p^2,\nn\\
\label{eqn:eom 4}
\frac{dp}{dt'}&=&\{p,H_\mubar^{(n)'}\}
=\frac{2}{\gamma}\cos(\mubar c)\,\mathfrak{S}_n(\mubar c)\,c_h^{(n)}p^2,
\ea
and the constraint that the Hamiltonian must vanish:
\be\label{eqn:eom 5}
H_\mubar^{(n)'}=0\quad\Rightarrow\quad
\frac{p_\phi^2}{2}=\frac{3}{8\pi G\gamma^2}\,(c_h^{(n)})^2p^2.
\ee
In particular, these lead to
\be\label{eqn:diff eq for p}
\frac{1}{p}\frac{dp}{d\phi}=\sqrt{\frac{16\pi G}{3}}\,
\cos(\mubar c)\,\mathfrak{S}_n(\mubar c),
\ee
as $\phi$ is treated as the internal time.

In the classical regime, we have $\mubar c\ll 1$ and thus $\cos(\mubar c)\rightarrow 1$, $\sin(\mubar c)\rightarrow 0$, $\mathfrak{S}_n(\mubar c)\rightarrow 1$ and $c_h^{(n)}\rightarrow c$; therefore, the above equations all reduce to their classical counterparts. In the backward evolution, the quantum corrections are more and more significant as $\mubar c$ becomes appreciable. Eventually, $p$ gets bounced at the epoch when $\cos(\mubar c)$ in \eqnref{eqn:diff eq for p} flips signs. The exact point of the quantum bounce is given by $\cos(\mubar c)=0$ (i.e. $\mubar c=\pi/2$). By \eqnref{eqn:holonomized c}, this happens when
\be
c_h^{(n)}\mubar=\sum_{k=0}^{n} \frac{(2k)!}{2^{2k}(k!)^2(2k+1)}
=:\mathfrak{F}_n
\ee
(note that $\mathfrak{F}_n\rightarrow \pi/2$ as $n\rightarrow\infty$), or equivalently, by \eqnref{eqn:eom 5}, when
\be
\label{eqn:rho crit}
\rho_\phi=\rho_\textrm{crit}^{(n)}:=3\mathfrak{F}_n^2\,\rho_\textrm{Pl}
\ee
if the improved scheme \eqnref{eqn:mubar scheme} is adopted.
Note that
\be
\label{eqn:rho crit infty}
3\rho_\textrm{Pl}=
\rho_\textrm{crit}^{(0)} < \rho_\textrm{crit}^{(1)} < \cdots < \rho_\textrm{crit}^{(\infty)}=\frac{3\pi^2}{4}\rho_\textrm{Pl}.
\ee

This shows that the occurrence of the quantum bounce and the boundedness of the matter density are generic for any given $n$. Even at the limit $n\rightarrow\infty$, the matter density remains bounded, as opposed to the classical theory.

\begin{widetext}

\begin{figure}
\begin{picture}(500,190)(0,0)

\put(-15,-15)
{
\scalebox{0.9}{\includegraphics{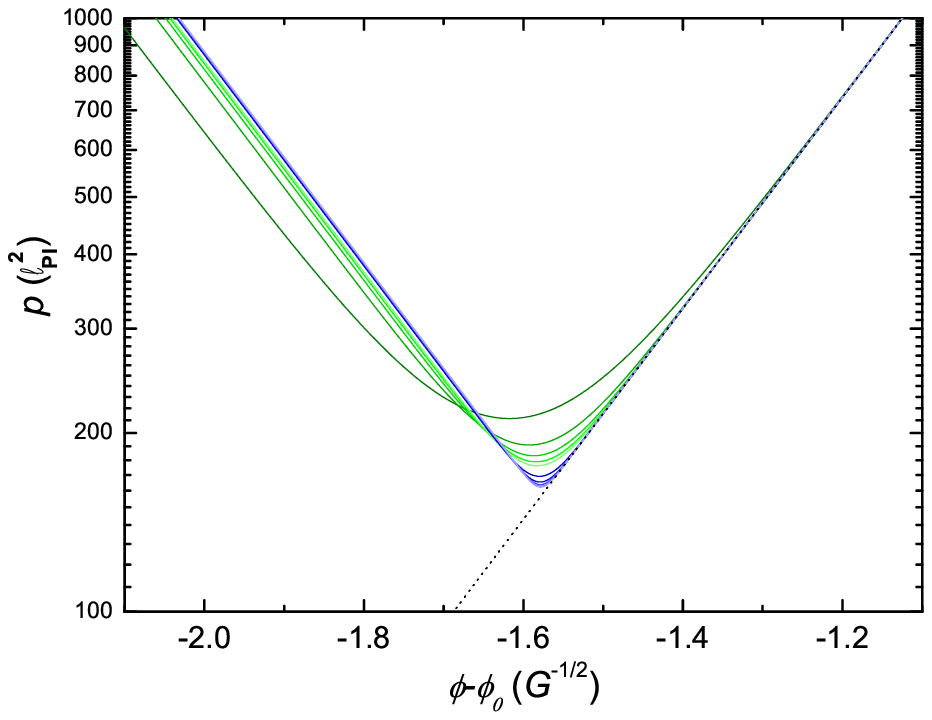}}
}

\put(250,-13.5)
{
\scalebox{0.9}{\includegraphics{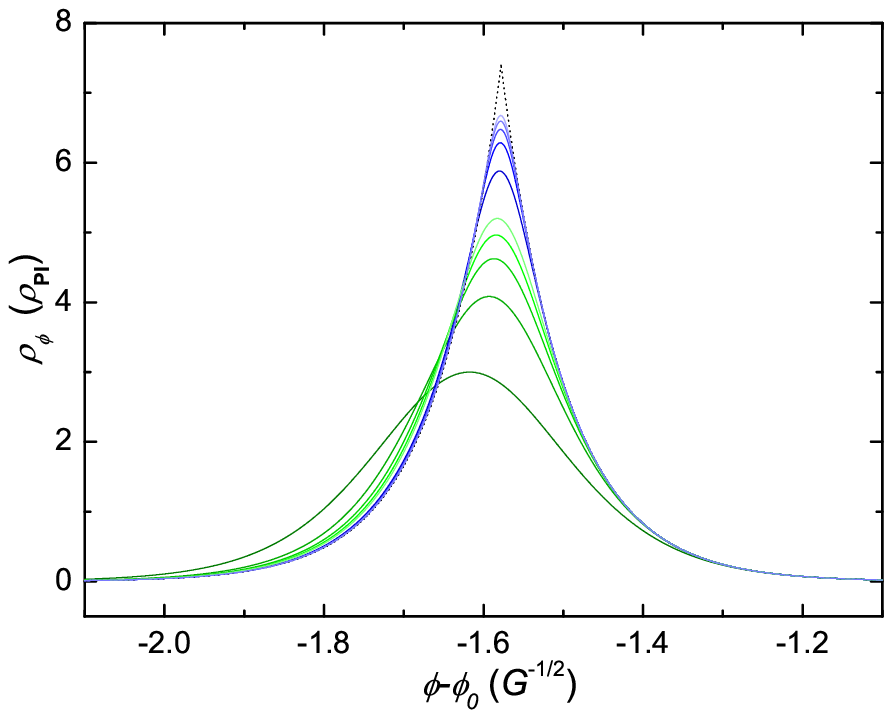}}
}

\put(0,180){\textbf{(a)}}
\put(260,180){\textbf{(b)}}

\end{picture}
\caption{Solutions of the heuristic effective dynamics with $p_\phi=2.\times10^3\hbar\sqrt{8\pi G}$, the initial value $p(\phi_0)=1.\times10^5\Pl^2$ and $\gamma=\ln2/(\sqrt{3}\pi)$. \textbf{(a)} Solid curves represent $p(\phi)$ for $n=0,1,2,3,4,10,20,30,40,50$ (the larger $n$ is, the smaller $p$ is at the bouncing point). The dotted line is the classical solution. \textbf{(b)} Solid curves are the corresponding $\rho_\phi$, the peaks of which agree with $\rho_\textrm{crit}^{(n)}$ given by \eqnref{eqn:rho crit} (the larger $n$ is, the bigger $\rho_\textrm{crit}^{(n)}$ is). The solution of $n=\infty$ is also indicated as the dotted curve, which is sharply kinked at the top tip at the value of $\rho_\textrm{crit}^{(\infty)}$ given by \eqnref{eqn:rho crit infty}.}\label{fig:numerical solutions}
\end{figure}

\end{widetext}

\section{Discussion}\label{sec:discussion}
For a given initial condition which satisfies the Hamiltonian constraint \eqnref{eqn:eom 5}, the differential equations \eqnref{eqn:eom 1}--\eqnref{eqn:eom 4} can be solved numerically (by the Runge-Kutta method) to yield the detailed evolution of the heuristic dynamics for a given $n$. The numerical solutions with \eqnref{eqn:mubar scheme} imposed are depicted in \figref{fig:numerical solutions} for different $n$.\footnote{The double precision used in the computer program loses necessary accuracy and results in noticeable round-off error when $n$ is huge ($n\agt100$); thus, the numerical computations are done only up to $n=50$.} It is affirmed that the nonsingular bouncing scenario is robust regardless of $n$: Two classical solutions (expanding and contracting) are bridged by the quantum bounce, which takes place when $\rho_\phi$ approaches the critical value $\rho_\textrm{crit}^{(n)}$.

As $n$ increases, the upper bound $\rho_\textrm{crit}^{(n)}$ increases as well, but the limiting value $\rho_\textrm{crit}^{(\infty)}$ remains finite in the order of the Planckian density as indicated in \eqnref{eqn:rho crit infty}. Furthermore, in \figref{fig:numerical solutions}(a) it is noted that the larger $n$ is, the longer the effective solution follows the classical trajectory before being deviated by the bounce. Put differently, the quantum bounce takes place more abruptly for a larger $n$. In the extreme case of $n\rightarrow\infty$, the effective solution exactly matches the classical solutions on both sides of the bounce, which is so abrupt that it only imprints a kink on the solution of $p(\phi)$. This is expected, since formally $c_h^{(n)}$ approximates $c$ closer and closer as $n$ increases, but nevertheless the factor on the right-hand side of \eqnref{eqn:diff eq for p} yields the limit: $\cos(\mubar c)\,\mathfrak{S_n}(\mubar c)\rightarrow\cos(\mubar c)\abs{\cos(\mubar c)}^{-1}=\textrm{sgn}(\cos(\mubar c))$, which gives rise to the quantum bounce as a kink. The remarkable point is that, even in the limit $n\rightarrow\infty$, the heuristic effective dynamics does not simply reduce to the classical one.

If the quantum theory of LQC with holonomy corrections up to the $n$th order can be constructed, following the lesson of \cite{Ashtekar:2007em}, we expect that the expectation value of the matter density remains finite as $\rho_\textrm{crit}^{(n)}$ obtained in \eqnref{eqn:rho crit} sets the absolute upper bound. Even if all orders of holonomy corrections are included, it is very likely that the quantum bounce persists and the expectation value of the matter density remains lower than $\rho_\textrm{crit}^{(\infty)}$ given by \eqnref{eqn:rho crit infty}. This suggests that, even with $n\rightarrow\infty$, the quantum theory of LQC, if it can be constructed, is fundamentally different from the WDW theory. Furthermore, in the quantum theory of LQC of $n\rightarrow\infty$, the abruptness of the bounce could be well smoothed by the quantum fluctuations.

As higher orders of holonomy corrections are included, $c_h^{(n)}$ gets closer to formally agreeing with $c$, yet the heuristic analysis shows that the matter density remains bounded and the bouncing scenario holds for any arbitrary $n$. This makes it more convincing to assert that the occurrence of the quantum bounce is essentially a consequence of the loopy nature (i.e. intrinsic discreteness) of LQC. The fact that it is well motivated and makes sense to incorporate higher order holonomy corrections at the level of heuristic effective dynamics encourages one to construct the quantum theory of LQC based on the new Hamiltonian \eqnref{eqn:H mubar n}, the investigation of which might in turn shed light on the issues of $j$ ambiguity for the Hamiltonian operator in the full theory of LQG.

The observation that the bouncing scenario survives the higher order holonomy corrections is also expected for other extended models. With the inclusion of generic matters, for example, the resulting effective solutions should be qualitatively the same as those of the familiar $n=0$ formalism, since the matter part of the Hamiltonian constraint involves only $p$ but no $c$ and hence receives no holonomy corrections. Consequently, the scenario in Appendix A of \cite{Ashtekar:2006wn} with a nonzero cosmological constant and those in \cite{Ashtekar:2006es,Vandersloot:2006ws} for the $k=\pm1$ FRW models should live on, because at the level of effective dynamics the cosmological constant and the spatial curvature terms can be treated as matter with the equation of state parameters $w=-1$ and $-1/3$, respectively. In the context of anisotropic models, the same is also anticipated, but the exact value of the absolute upper bound for the matter density is difficult to pinpoint, as the effective equations of motion are more complicated and the exact condition for the bounce depends on the degree of anisotropy (see \cite{Chiou:2007mg} for the Bianchi I model and \cite{Chiou:2008eg} for the Kantowski-Sachs spacetime for the case of $n=0$). Similarly, the loop quantum geometry of the Schwarzschild black hole interior predicted in \cite{Chiou:2008nm} is not expected to be spoiled either, but again the details demand closer examination.

Additionally, as remarked earlier in \secref{sec:holonomy corrections}, it should be noted that the extension of higher order holonomy corrections studied in this paper is distinct from that in \cite{Mielczarek:2008zz,Hrycyna:2008yu}. Opposed to our conclusion, \cite{Mielczarek:2008zz} and \cite{Hrycyna:2008yu} suggest that the nonsingular bouncing scenario that appears in the lowest order could only be an artifact of simplification, since it is qualitatively modified by the $\mathcal{O}(\mubar^4)$ corrections. As it is debatable which extension approach makes more sense, whether the higher order holonomy corrections modify the bouncing scenario qualitatively or only quantitatively remains an open question. To give a decisive answer, construction of the quantum theory of LQC with the inclusion of higher order holonomy corrections and further investigations from the perspective of LQG are necessary.

\newpage

\begin{acknowledgements}
The authors would like to thank Jakub Mielczarek for bringing the related works to their attention and giving valuable comments.
DWC is supported by the NSFC Grant No. 10675019 and LFL is supported by the NSFC Grant No. 10875012.
\end{acknowledgements}


\end{document}